\documentclass{ws-ijmpa}

\begin{document}

\markboth{A. Bonanno, G. Esposito, C. Rubano}
{Improved Action Functionals in Non-Perturbative Quantum Gravity}

\catchline{}{}{}

\title{IMPROVED ACTION FUNCTIONALS IN NON-PERTURBATIVE QUANTUM GRAVITY}

\author{A. BONANNO,$^{1,2}$ G. ESPOSITO,$^{3,4}$ C. RUBANO$^{4,3}$}

\address{${ }^{1}$Osservatorio Astrofisico, Via S. Sofia 78, 
95123 Catania, Italy\\
${ }^{2}$INFN, Sezione di Catania,
Corso Italia 57, 95129 Catania, Italy\\
${ }^{3}$INFN, Sezione di Napoli,
Complesso Universitario di Monte S. Angelo,\\ Via Cintia, Edificio N',
80126 Napoli, Italy\\
${ }^{4}$Dipartimento di Scienze Fisiche, Universit\`a Federico II,
Complesso Universitario\\ di Monte S. Angelo, Via Cintia, Edificio N',
80126 Napoli, Italy\\
Received 2 September 2004}

\maketitle

\begin{abstract}
Models of gravity with variable $G$ and $\Lambda$ have acquired
greater relevance after the recent evidence in favour of
the Einstein theory being non-perturbatively 
renormalizable in the Weinberg sense. The present paper builds
a modified Arnowitt--Deser--Misner (ADM) action functional
for such models which
leads to a power-law growth of the scale factor
for pure gravity and for a massless $\phi^4$ theory in a Universe
with Robertson--Walker symmetry, 
in agreement with the recently developed 
fixed-point cosmology. Interestingly, the renormalization-group flow
at the fixed point is found to be compatible with a Lagrangian
description of the running quantities $G$ and $\Lambda$.
\end{abstract}

\section{Introduction}	

The main idea of the Renormalization Group (hereafter RG) 
approach to field theory is to
``integrate out'' all fluctuations with momenta larger than
some cutoff $k$, and to take them into account by means of a
modified dynamics for the remaining fluctuation modes with 
momenta smaller than $k$. This modified dynamics is governed
by a scale-dependent effective action $\Gamma_{k}$, the
$k$-dependence of which is described by a functional differential
equation, the exact RG equation.

Flow equations can be used for a complete quantization of fundamental
theories. If the latter have the classical action $S$ one imposes the
initial condition $\Gamma_{\kappa}=S$ at the ultraviolet (UV) cut-off
scale $\kappa$, uses the exact RG equation to compute $\Gamma_{k}$
for all $k < \kappa$, and then sends $k \rightarrow 0$ and
$\kappa \rightarrow \infty$. The defining property of a fundamental 
theory is that the ``continuum limit'' $\kappa \rightarrow \infty$ 
actually exists after the renormalization of finitely many parameters
in the action. 
Interestingly, there are also perturbatively non-renormalizable 
theories which admit a $\kappa \rightarrow \infty$ limit. The 
``continuum limit'' of these 
theories is taken at a non-Gaussian fixed point of the RG flow.
It replaces the Gaussian fixed point which, at least implicitly,
underlies the construction of perturbatively renormalizable theories.
Recently, Lauscher and Reuter$^{1}$ have constructed a new exact RG
equation for the effective average action of Euclidean quantum
gravity. They have found both a Gaussian and a non-Gaussian fixed
point. A strong evidence has been
therefore obtained in
favour of 4-dimensional Einstein gravity being asymptotically
safe.$^{2}$ 
Here, the key question we study is {\it how to achieve a modification
of the standard ADM Lagrangian where $G$ and $\Lambda$ are dynamical
variables according to a prescribed renormalized trajectory}.
Moreover, we are interested in the cosmological applications of
such a framework.$^{3}$ 

\section{Modified Action Functional}

We here use the ADM Hamiltonian treatment of 
space-time geometry. Thus, the identity 
\begin{equation}
\sqrt{-g} \; { }^{(4)}R=N \sqrt{h}(K_{ij}K^{ij}-K^{2}
+{ }^{(3)}R)-2(K \sqrt{h})_{,0}+2f_{\; ,i}^{i},
\label{(1)}
\end{equation}
where 
$
f^{i} \equiv \sqrt{h}\Bigr(KN^{i}-h^{ij}N_{,j}\Bigr)
$,
suggests using the Leibniz rule to express
\begin{equation}
{1\over G}(K \sqrt{h})_{,0}
={G_{,0}\over G^{2}}K \sqrt{h}+
\left({K \sqrt{h}\over G}\right)_{,0},
{1\over G}{\partial f^{i}\over \partial x^{i}}
={G_{,i}\over G^{2}}f^{i}+{\partial \over \partial x^{i}}
\left({f^{i}\over G}\right),
\label{(2)}
\end{equation}
so that division by $16 \pi G$ in the integrand of the
Einstein-Hilbert action yields the Lagrangian (the $c^{4}$
factor in the numerator is set to $1$ with our choice of units)
\begin{equation}
L={1\over 16\pi}\int \left[{N \sqrt{h}\over G}(K_{ij}K^{ij}
-K^{2}+{ }^{(3)}R-2 \Lambda)
-2{G_{,0}\over G^{2}}K \sqrt{h}
+2{G_{,i}f^{i}\over G^{2}}\right]d^{3}x,
\label{(3)}
\end{equation}
after adding to the action functional the boundary term
$
I_{\Sigma}={1\over 8\pi}\int_{\Sigma}{K \sqrt{h}\over G}d^{3}x
$,
and assuming that $\Sigma$ is a closed manifold. 
The Lagrangian (3), however, suffers from a serious drawback,
because the resulting momentum conjugate to the three-metric reads as
\begin{equation}
p^{ij} \equiv {\delta L \over \delta h_{ij,0}}
=-{\sqrt{h}\over 16 \pi G}(K^{ij}-h^{ij}K)
+{\sqrt{h}\over 16 \pi G}{h^{ij}\over NG}(G_{,0}-G_{,k}N^{k}).
\label{(4)}
\end{equation}
This would yield, in turn, a Hamiltonian containing a term quadratic
in $G_{,0}$ (since $K^{ij}$ depends also on 
$G_{,0}h^{ij}$), despite the fact that (3) is only linear in
$G_{,0}$ when expressed in terms of first and second fundamental
forms. There is therefore a worrying lack of equivalence between $K^{ij}$
and $p^{ij}$. 
We thus decide to include 
a ``bulk'' contribution in order to cancel the effect of
$G_{,0}$ and $G_{,i}$ in Eq. (3) by writing 
\begin{equation}
I \equiv {1\over 16\pi}\int_{M}{({ }^{(4)}R-2\Lambda)\over 
G({\vec x},t)}\sqrt{-g} \; d^{4}x
+{1\over 8\pi}\int_{M}{(K \sqrt{h})_{,0}\over 
G({\vec x},t)}d^{4}x
-{1\over 8\pi}\int_{M}{f_{\; ,i}^{i}\over 
G({\vec x},t)}d^{4}x,
\label{(5)}
\end{equation}
as a starting action defining a gravitational theory with variable $G$ and 
$\Lambda$. More precisely, upon division by $16 \pi G$ in the integrand
of the Einstein--Hilbert term (first integral in Eq. (5)), 
the second and third integral in Eq. (5)
cancel the effect of $-2(K\sqrt{h})_{,0}$ and $2f_{\; ,i}^{i}$ in Eq.
(1), respectively. The resulting Lagrangian density belongs to the
general family depending only on fields and their first derivatives,
which is the standard assumption in local field theory.
It should also be noticed that if $G$ were constant, 
the second integral on the right-hand side
of (5) would reduce to the York--Gibbons--Hawking boundary term 
$
{1\over 8\pi G}\int_{\Sigma}K \sqrt{h} \; d^{3}x
$.

In other words, we are trying to 
generalize the standard ADM Lagrangian in order to consider $G$ as a 
{\it dynamical field} and investigate which dynamics 
is consistent with the RG approach. In this
spirit we consider hereafter the following general ADM
Lagrangian:
\begin{equation}
L={1\over 16\pi}\int \left[{N \sqrt{h}\over G}(K_{ij}K^{ij}
-K^{2}+{ }^{(3)}R-2 \Lambda)\right]d^{3}x +L_{\rm int}+L_{\rm matter},
\label{(6)}
\end{equation}
where the first term has the same functional form as the Lagrangian
of ADM general relativity (but with $G$ and $\Lambda$ promoted to
dynamical variables),
$L_{\rm int}$ is an interaction term of a kinetic type 
which, for dimensional reasons, must be of the form
(the coefficient $16 \pi$ is introduced for later convenience)
\begin{equation}
L_{\rm int}=  -{\mu \over 16 \pi}
\int  \frac{g^{\rho \sigma}G_{;\rho}G_{;\sigma}}{G^3}\; 
\sqrt{-g} \; d^{3}x,
\label{(7)}
\end{equation}
$\mu$ being the interaction parameter.
$L_{\rm matter}$ is the ``matter'' Lagrangian that
we shall consider as given by a self-interacting scalar field.

\section{RW Symmetry}

On using the ADM formalism, we take a 
scalar-field Lagrangian 
\begin{equation}
L_{m} \equiv \int {N \sqrt{h}\over 2}\left[N^{-2}(\phi_{,0})^{2}
-2{N^{i}\over N^{2}}\phi_{,0}\phi_{,i}
-\left(h^{ij}-{N^{i}N^{j}\over N^{2}}\right)\phi_{,i}\phi_{,j}
-2V(\phi) \right]d^{3}x,
\label{(8)}
\end{equation}
where the potential $V(\phi)$ is, for the time being, un-determined,
and $g^{00}$ is negative with our convention for the space-time
metric. 

We focus, hereafter, on cosmological models with RW symmetry.
Strictly speaking, such a name can be criticized, since we are no
longer studying general relativity, nor are we simply RG-improving
the Einstein equations. Nevertheless, we will find that spatially
homogeneous and isotropic cosmological models of the RW class can
still be achieved. In such models with lapse function $N=1$, 
the full Lagrangian, including scalar field, reads as (here 
${\cal K}=1,0,-1$ for a closed, spatially flat or open universe,
respectively)
\begin{equation}
L={a^{3}\over 16 \pi G}\left(-6{{\dot a}^{2}\over a^{2}}
+{6{\cal K}\over a^{2}}-2\Lambda \right)
+ {\mu \over 16 \pi} {a^{3}{\dot G}^{2}\over G^3}
+a^{3}\left({{\dot \phi}^{2}\over 2}-V(\phi)\right),
\label{(9)}
\end{equation}
where hereafter we revert to dots, for simplicity, to denote
derivatives with respect to $t$. Thus, the resulting second-order
evolution equations for $a,G$ and $\phi$ are 
\begin{eqnarray}
&&{{\ddot a} \over a}
+{1\over 2} {{\dot a}^{2} \over a^{2}}
+{{\cal K} \over 2 a^{2}}-{\Lambda \over 2}
-{{\dot a} \over a}{{\dot G} \over G}+{\mu\over 4}  
{{\dot G}^2 \over G^2} +
4 \pi G \left({{\dot \phi}^{2}\over 2}-V(\phi)\right)=0, \\[2mm]
&&
\mu{\ddot G}
-{3\over 2}\mu {{\dot G}^2\over G}
+3\mu{\dot a \over a} {\dot G}  
+{G \over 2 }\left(-6{{\dot a}^{2}\over a^{2}}
+6{{\cal K} \over a^{2}}-2\Lambda 
+2G{d \Lambda \over d G}\right)=0, \label{eqg}\\[2mm]
&&{\ddot \phi}+3{{\dot a}\over a}{\dot \phi}
+{dV \over d\phi}=0. 
\end{eqnarray}
Moreover, since the Lagrangian (6) is independent of time derivatives of
the lapse, one has the primary constraint of vanishing conjugate
momentum to $N$. The preservation in time of such a primary constraint
yields, for our Lagrangian generated from the assumption of RW symmetry,
the constraint equation
\begin{equation}
{{\dot a}^{2}\over a^{2}}+{{\cal K} \over a^{2}}
-{\Lambda \over 3}-{\mu \over 6}  {{\dot G}^{2} \over G^2}
-{8\pi G \over 3}\left({{\dot \phi}^{2}\over 2}+V(\phi)\right)
= 0 .
\label{(13)}
\end{equation}

In cosmology, the postulate of homogeneity and
isotropy implies that the RG scale dependence is
turned into a time dependence:
$
G(t)\equiv G(k=k(t)), \Lambda(t) \equiv \Lambda(k=k(t))
$. The explicit law is
$
k(t) = {\xi} /{t}
$,
$\xi$ being a positive constant, because, 
when the age of the Universe is $t$, 
no fluctuation with a frequency
smaller than $1/t$ can have played any role as yet. 
Hence the integrating-out of modes which underlies the Wilson
renormalization group should be stopped 
at $k\approx 1/t$. In the neighbourhood of a fixed point 
$(g_\star,\lambda_\star)$
the evolution of the dimensionful $G$ and 
$\Lambda$ is approximately given by
$
G(k) = \frac{g_{\star}}{k^2}, \Lambda(k) 
= \lambda_{\star} \; k^{2}
$. 
We therefore obtain the time-dependent 
Newton parameter and cosmological term:
\begin{equation}
G(t) = g_\star \xi^{-2} \; t^{2},\;\;\;\; \Lambda(t) 
= \frac{\lambda_\star \xi^2}{t^2}.
\label{(14)}
\end{equation}

Unlike models where only the Einstein equations are RG-improved,
our framework allows for a non-trivial dynamics of the scale factor
even in the absence of coupling to an external field.$^{4}$
To appreciate this point, consider first the case
when no scalar field exists, so that Eq. (12) 
should not be considered. 
The relation (14) suggests looking for power-law solutions of the type
\begin{equation}
a(t)=A \;t^\alpha, 
\;\;\;\;\; G(t)=  g_\star \xi^{-2} t^{2},\;\;\;\; 
\Lambda(t)= \lambda_\star \xi^2 t^{-2},
\label{(15)}
\end{equation}
and separately consider the ${\cal K}=0$ and ${\cal K}=\pm 1$ case. 
For ${\cal K}=0$ we obtain that $A$ is an un-determined factor, while 
(from Eq. (13))
\begin{equation}
\mu_\pm = \frac{1}{4} (3\pm\sqrt{9+12 \;\xi^2\lambda_\star}) , \;\;\;\; 
\alpha_\pm = \frac{1}{6}(3\pm\sqrt{9+12\;\xi^2\lambda_\star}),
\label{(16)}
\end{equation}
which implies a power-law inflation for the ``$+$'' solution, 
$\alpha_+$ being larger than $1$ if $\lambda_\star>0$, 
and a possible solution of the horizon problem. 
Note that the first equality is 
a relation between coupling constants which has to be 
satisfied, while the second simply relates the
value of $\alpha$ with the product $\xi^2 \lambda_\star$. 
Since $\xi$ is not determined,
$\alpha_+$ can be made arbitrarily large. Moreover, 
both $\Omega_\mu$ and $\Omega_\Lambda$
are constant, since 
\begin{equation}
\Omega_\mu = 1-\Omega_\Lambda = \frac{6}{3\pm\sqrt{9+12 \; 
\xi^2\lambda_\star}}.
\label{(17)}
\end{equation}

If ${\cal K}=\pm 1$ we find instead $\alpha =1$ and 
\begin{equation}
\mu = \frac{1}{4}(6+\xi^2\lambda_\star) , \;\;\;\;\;\;\; 
\frac{{\cal K}}{A^2} = \frac{\xi^2\lambda_\star}{2},
\label{(18)}
\end{equation}
where, as before, the former equation relates the values of 
the coupling constants, while the latter 
is a consistency relation. In particular we see that, if 
${\cal K}=-1$, then $\lambda_\star$ must be negative.
In both cases (i.e. ${\cal K}=\pm 1$),  
$\Omega_\Lambda$ and $\Omega_\mu$ are constant with 
\begin{equation}
\Omega_\Lambda = \frac{2{\cal K}}{3 A^2}, \;\;\;\;\; 
\Omega_\mu = 1+\frac{{\cal K}}{3A^2}.
\label{(19)}
\end{equation}

Next we consider the contribution of a scalar field with 
a self-interacting potential of the type 
$
V(\phi) = \frac{\zeta}{4!}\phi^{4}
$,
with the ansatz (15) for $a,G,\Lambda$,  
and $\phi = \phi_0 t^{-\beta}$ for the scalar field. The
Klein--Gordon equation of motion (12) then yields $\beta=1$ and
$\phi_0= \pm \sqrt{\frac{6(3\alpha -2)}{\zeta}}$,
which implies $\alpha>2/3$ so as to have a real scalar field. 
We then find, if ${\cal K}=0$,
\begin{eqnarray}
\label{(20)}
&&\alpha_{\pm} = \mu \pm \sqrt{\mu^2-\frac{4\mu}{3}
-\frac{2\xi^2\lambda_\star}{3} }, \\[2mm]
&&\zeta_\pm = \frac{ 
24 g_\star\pi[3\xi^2\lambda_\star-3\mu^{2}+5\mu 
\pm (1-\mu)\sqrt{3\mu (3\mu-4)-6\xi^2\lambda_\star}]} 
{\xi^2(3\xi^2\lambda_\star +2(3-2\mu)\mu)},
\label{(21)}
\end{eqnarray}
where (21) is a consistency relation, 
(20) determines the value of $\alpha$, and a 
reality condition for $\xi^2\lambda_\star$ is given by 
$(3\xi^2\lambda_\star +2(3-2\mu)\mu)>0$. 
It is not difficult to see that there are physically interesting 
solutions with power-law inflation if
$\mu$ is large enough and positive, and $\xi$ is positive. 
If instead ${\cal K}=\pm 1$ the only solution is for 
$\alpha =1, \beta=1,$ 
and we get the consistency conditions$^{4}$
\begin{equation}
\mu = - \frac{9 \pi \xi^{-2} g_\star}{\zeta} +\frac{1}{4}
(6+\xi^2 \lambda_\star),  \;\;\;\;\;\;
\frac{\cal K}{A^2} = \frac{6 \pi \xi^{-2} g_\star}{\zeta}
+ {\xi^{2}\lambda_{\star}\over 2}, 
\;\;\;\;\; \phi_0 = \pm \sqrt {\frac{6}{\zeta}}. 
\label{(22)}
\end{equation}

A challenging open problem is now the development of cosmological
perturbation theory starting from Lagrangians as in Eq. (6). This
could tell us whether formation of structure in the early universe
can also be accommodated within the framework of a Wilson-type$^{5}$
formulation of quantum gravity.

\section*{Acknowledgments}

The authors are grateful to the INFN for financial support. The
work of G. Esposito has been partially supported by PRIN 2002
``Sintesi''.

\end{document}